\documentclass{ws-procs9x6}

\begin{document}

\title{The wedding of modified dynamics \\ and non-exotic dark matter in galaxy clusters}

\author{B. FAMAEY$^{1,*}$, G. W. ANGUS$^2$, G. GENTILE$^3$, H. Y. SHAN$^4$, H. S. ZHAO$^{2,4}$}

\address{$^1$IAA, Universit\'e
Libre de Bruxelles,
Bvd du Triomphe, 1050 Bruxelles, Belgium \\
$^2$SUPA, Univ. of St. Andrews, KY16 9SS Fife, UK \\
$^3$Univ. of New Mexico, 800 Yale Blvd NE, Albuquerque, New Mexico 87131, USA\\
$^4$National Astronomical Observatories, Beijing 100012, PRC\\
$^*$E-mail: bfamaey@ulb.ac.be}

\begin{abstract}
We summarize the status of Modified Newtonian Dynamics (MOND) in galaxy clusters. The observed acceleration is typically larger than the acceleration threshold of MOND in the central regions, implying that some dark matter is necessary to explain the mass discrepancy there. A plausible resolution of this issue is that the unseen mass in MOND is in the form of ordinary neutrinos with masses just below the experimentally detectable limit. In particular, we show that the lensing mass reconstructions of the rich clusters 1E0657-56 (the bullet cluster) and Cl0024+17 (the ring) do not pose a new challenge to this scenario. However, the mass discrepancy for cool X-ray emitting groups in which neutrinos cannot cluster pose a more serious problem, meaning that dark baryons could present a more satisfactory solution to the problem of unseen mass in MOND clusters.
\end{abstract}


\bodymatter

\section{Introduction}

Data on large scale structures point towards a Universe dominated by dark matter and dark energy \cite{sperg}. Discovering the nature of these mysterious components of the Universe is, without a 
doubt, the major challenge of modern astrophysics, nay of physics as a whole. Nowadays, the dominant paradigm is that dark matter is actually made of non-baryonic weakly interacting massive particles, the so-called ``cold dark matter'' (CDM), and that the mysterious dark energy is well represented by a cosmological constant ($\Lambda$) in Einstein equations. The $\Lambda$CDM cosmological model has known a remarkable success in explaining and predicting diverse data sets corresponding to the Universe at its largest scales, including the CMB radiation, galaxy redshift surveys, distant supernovae data and absorption lines in the spectra of distant quasars. Nevertheless, a number of observations on galactic scales appear to be at variance with a number of 
CDM predictions. For instance, measurements of non-circular motions in the
Milky Way have shown that there is actually very little room for dark matter inside the solar radius \cite{FB05}, where CDM simulations predict a cuspy density profile. External galaxies have also been used to compare the predicted cuspy CDM density profiles with the observations, in particular rotation curves of dwarf and spiral galaxies show evidence for dark matter halos with a central constant density core \cite{Gen04} at odds with the CDM predictions. Another interesting problem faced by CDM on galactic scales is the overabundance of predicted satellite galaxies compared to the observed number in Milky Way-sized galaxies \cite{M99}. What is more, it is now well-documented that rotation curves suggest a correlation between the mass profiles of the baryonic matter (stars + gas) and dark matter \cite{Don04}. Some rotation curves, like the one of NGC1560 \cite{Bro92} even display obvious features (bumps  or wiggles) that  are also clearly  visible in  the stellar or gas distribution. A solution to all these problems, and especially the baryon-DM relation, could be a new specific interaction between baryons and some exotic dark matter made of, e.g., dipolar particles \cite{Bla07, F07}. On the other hand, it could indicate that, on galaxy scales, the observed discrepancy rather reflects a breakdown of Newtonian dynamics in the ultra-weak field regime: this alternative explanation to solve the dark matter problem is known as the Modified Newtonian Dynamics (MOND\cite{Mil83}) paradigm, which postulates that for accelerations below $a_0 \approx 10^{-10} {\rm m} \, {\rm s}^{-2}$ the effective gravitational attraction approaches $(g_N a_0)^{1/2}$ where $g_N$ is the usual Newtonian gravitational field. Without resorting to galactic dark matter, this simple prescription is known to reproduce galaxy scaling relations in spirals and ellipticals (Tully-Fisher, Faber-Jackson, fundamental plane) as well as the details of the rotation curves of individual spiral galaxies \cite{SM02} over five decades in mass. In particular, the recent kinematic analysis of tidal dwarf galaxies belonging to the NGC~5291 system \cite{Bo07}, showing a mass discrepancy unexpected in the CDM context, strongly argues in favour of MOND\cite{Gen07, Mil07}. Moreover, the paradigm successfully predicts the local galactic escape speed from the solar neighbourhood \cite{FBZ07,W07}, the statistical bar frequency in spirals\cite{TC07}, as well as the velocity 
dispersions of satellite galaxies around their hosts\cite{KP1,KP2}. Recent developments in the theory of gravity have also added plausibility to the case for modification of gravity through the advent of Lorentz-covariant theories of gravity yielding a MOND behaviour in the appropriate limit\cite{Bek04, Zlos1, BEF}. Although rather fine-tuned and still being a far cry from a fundamental theory underpinning the MOND paradigm, these theories remarkably allow for new predictions regarding cosmology \cite{Sk06,DL06,Zlos2,HZ} and gravitational lensing\cite{Z06,Xu07}. Hereafter we notably investigate the weak-lensing properties of some galaxy clusters in MOND.

\section{The modified dynamics in galaxy clusters}

While having an amazing predictive power on galactic scales, the simple MOND prescription badly fails in galaxy clusters without an additional unseen component. Indeed, in rich clusters of galaxies, the observed acceleration is typically larger than $a_0$ in the central regions, meaning that the MOND prescription is not enough to explain the observed discrepancy between visible and dynamical mass there\cite{A01,S03,S07}, a conclusion that can be reached by computing the centripetal gravity as a function of radius in the cluster (and thus the corresponding enclosed MOND mass) from the density and temperature profiles of X-ray gas and from the assumed hydrostatic equilibrium of the cluster.

At very large radii, the discrepancy is about a factor of two, meaning that there should be as much dark matter (mainly in the central parts) as observed baryons in MOND clusters. The main characteristic of this MOND dark matter is thus that it should cluster at galaxy cluster scales but not at galaxy scales. An ideal candidate, whose free-streaming length is known to be high, is at the same time the only dark matter particle that we know for sure to exist, the neutrino. We know that ordinary neutrinos have mass\cite{Fuk98} and that they have a number density comparable to photons, meaning that they indeed contribute to the mass budget of the Universe. However, in order to reach the densities needed to account for the MOND missing mass in galaxy clusters, they should have a mass at the limit of their experimental detection, i.e. 2~eV. This idea\cite{S03} has the great advantage of naturally reproducing most cluster scaling relations including the luminosity-temperature relation\cite{S07}, while accounting for the bulk of the missing mass in galaxy clusters. Moreover, in their modelling of the CMB anisotropies, Skordis et al.\cite{Sk06} showed that such a significant non-baryonic component (with $\Omega_n \simeq 0.15$) was actually helpful to prevent the MOND Universe from accelerating too much, keeping $\Omega=1$ as a constraint on the amount of dark energy (although MOND might have the ability to drive late-time acceleration without resorting to dark energy\cite{Diaz}).

On the other hand, given that, in the global baryon inventory at low redshift, about 20\% of the baryons are still missing, and that the observed baryons in clusters only account for 5 to 10\% of those produced during Big Bang nucleosynthesis\cite{Fukugita,Silk,McG07}, there is plenty of room for this dark matter to be baryonic in MOND, since there should be as much dark matter (mainly in the central parts) as observed baryons in MOND clusters. Knowing exactly how many baryons hide in the Warm-Hot Intergalactic Medium (WHIM) is thus imperative if one wants to exclude this hypothesis.

\section{The bullet cluster 1E0657-56}

Keeping in mind this known discrepancy between the observable and dynamical masses of galaxy clusters in MOND, it is then useful to ask which new challenge is posed to the MOND paradigm by the gravitational lensing map of the bullet cluster\cite{Clowe,Bradac} (see M.~Bradac's contribution to these proceedings). In this extremely interesting object, the collisionless component (galaxies and a hypothesised collisionless dark matter component) and the fluid-like X-ray emitting plasma have been spatially segregated due to the collision of the two progenitor galaxy clusters. However, the lensing convergence map is centered on the minor baryonic collisionless component (galaxies) rather than on the dominant baryonic X-ray emitting gas component: this was argued \cite{Clowe} to be the first direct empirical proof of the existence of dark matter, independently of the validity of General Relativity at galaxy cluster scales. However, while the linear relation between the matter density and the gravitational potential implies that the convergence parameter is a direct measurement of the projected surface density in General Relativity, this is not the case anymore in MOND due to the non-linearity of the modified Poisson equation. Actually, it has been shown that, in MOND, it is possible to have a non-zero convergence along a line of sight where there is zero projected matter\cite{AFZ}. However, in the specific case of the bullet cluster, solving the non-linear Poisson equation for the observed matter density in various line-of-sight configurations showed that the convergence map always tracks the dominant baryonic component\cite{Feix}: this means that non-linear effects, being capable of counteracting this trend, turn out to be very small. The presence of large amounts of collisionless dark matter in this cluster is thus necessary in MOND. 

However, by applying a simple potential-density approach, we\cite{bullet} have been able to estimate the needed quantities of such collisionless dark matter in the bullet cluster, finding that the central densities around the galaxies were in accordance with the maximum density of 2~eV neutrinos, from the Tremaine-Gunn\cite{TG} limit for a 9~keV ($\sim 10^8$~K) cluster:
\begin{equation}
\rho_{\nu}^{\rm max}=7\times T({\rm keV})^{3/2} \times 10^{-5} {\rm M}_\odot {\rm pc}^{-3} 
\end{equation}
$\sim 2 \times 10^{-3} {\rm M}_\odot/{\rm pc}^3$. However, a problem might exist from strong lensing data at the center of the collisionless component of the least massive cluster, a problem similar to the one discussed in section 5. We however conclude that the weak-lensing map of the bullet cluster in itself is not a new challenge to the ``MOND+neutrinos" hypothesis,  meaning that the amount of dark matter required is globally consistent with that suggested by the previous analyses\cite{S03} from hydrostatic equilibrium of X-ray emitting clusters. However, if it turns out that the MOND dark matter should rather be in baryonic form, then the bullet cluster provided the interesting constraint that it should be of collisionless nature (e.g. MACHO's or dense clumps of cold gas, but see also Mahdavi et al.\cite{Mah} for a counter-example).

We finally note that possible non-trivial contributions from the vector field of relativistic MOND theories in non-stationary configurations \cite{DL06,Zlos2,HZ} were neglected, which could only decrease the need for dark matter in this system (but not in other clusters close to a steady-state equilibrium), and that the high-speed encounter of the clusters making up the bullet could actually be a standard manifestation of MOND long-range interaction\cite{AM07}.

\section{The ring in Cl0024+17}

Recently, a comprehensive weak lensing mass reconstruction of the rich galaxy cluster Cl0024+17 at $z=0.4$\cite{J07} has been argued to have revealed the first dark matter structure that is offset from both the gas and galaxies in the cluster. This structure is ringlike, located between $r \sim 60''$ and $r \sim 85''$. It was argued to be the result of a collision along the line-of-sight of two massive clusters 1-2 Gyr in the past. It has also been argued\cite{J07} that this offset was hard to explain in MOND. 

Assuming that this ringlike structure is real and not caused by instrumental bias or spurious effects in the weak lensing analysis (due e.g. to the unification of strong and weak-lensing), and that cluster stars and galaxies do not make up a high fraction of the mass in the ring (which would be too faint to observe anyway), is this really hard to explain in MOND? 

First of all, it has recently been shown \cite{MilSan} that, considering the boost of the gravitational field in MOND as the effect of some virtual dark matter (which makes it easier to compare with Newtonian and General Relativistic predictions), a peak in this virtual matter distribution {\it generically} appears close to the transition radius of MOND $r_t=(GM/a_0)^{1/2}$, especially when most of the mass of the system is well-contained inside this radius (which is the case for the cluster Cl0024+17). This means that the ring in  Cl0024+17 could be the first manifestation of this pure MOND phenomenon. However, the sharpness of this virtual dark matter peak strongly depends on the choice of the $\mu$-function, controlling the transition from the $1/r^2$ Newtonian regime to the $1/r$ MOND regime\cite{F07}. A sharp transition of the $\mu$-function is needed to reproduce the ringlike structure observed in Cl0024+17, meaning that if the simple $\mu$-function\cite{FB05,ZF} recently used to fit many galaxy rotation curves is chosen, the ring cannot be adequately reproduced by this pure MOND phenomenon.

In this case, a collisional scenario would be needed in MOND too, in order to explain the feature as a peak of cluster dark matter. Indeed, as explained above, we already know that there is a mass discrepancy in MOND clusters, and we know that this dark matter must be in collisionless from (e.g., neutrinos or dense clumps of cold gas). So the results of the simulation with purely collisionless dark particles\cite{J07} would surely be very similar in MOND gravity. In case the missing mass in clusters is in baryonic form, we do not really have a quantitative limit on the density of MOND dark matter that would be allowed in the ring. But since we know that the ``MOND + neutrinos" hypothesis works fine in other similar rich clusters, we can follow the approach of Angus et al.\cite{bullet} and test this hypothesis in Cl0024+17. If the missing mass is in the form of dark baryons, this is an effective way to compare the dark density to what should be expected in similar clusters in MOND. 

Let us note that this cluster was already studied \cite{Tak07} in the framework of MOND, however this was prior to the detection of the ringlike structure. The cluster was found to be marginally consistent with 2~eV neutrinos, using a Hernquist profile with a total mass of $3.5 \times 10^{14} {\rm M}_\odot$ and a core radius of 0.3~Mpc. In a latter version, a cored model was tried, including also 
the strong lensing data, and a model consistent with a neutrino mass of 4~eV was found. However, they assumed a simple spherical model without any line-of-sight structure, contrary to the spirit of the collision scenario invoked to explain the ringlike feature. Given the uncertainty of the density models, it is unclear if existing data for this system actually rule out the 2~eV neutrinos. We hereafter rather focus on the newly discovered ringlike structure to see if it presents a new challenge to the ``MOND+neutrinos" hypothesis.

The main limit on the neutrino ability to collapse in clusters comes from the Tremaine-Gunn limit\cite{TG}, stating that the phase space density must be preserved during collapse. Assuming the same temperature for the neutrino fluid as for the baryons, the maximum density of a mixture of all neutrino types all having a 2~eV mass for a cluster of a given temperature $T$ (in keV) is then given by Eq.~(1). This means that for Cl0024+17 whose mean emission weighted temperature is $T=4.25_{-0.35}^{+0.40}$~keV \cite{J07}, the Tremaine-Gunn limit for the density of neutrinos is $\rho_{\nu}^{\rm max}=6.1_{-0.7}^{+0.9} \times 10^{-4} {\rm M}_\odot {\rm pc}^{-3}$.

A detailed simulation of Cl0024+17 would involve numerically solving the non-linear Poisson equation of MOND. However since observationally consistent relativistic MOND theories\cite{Bek04,Zlos1} always enhance the gravitational lensing, the surface density of the ring derived from General Relativity is always an upper limit to the actual density in MOND. Moreover, the gravity at the position of the ring is of the order of $\sim 2a_0$, meaning that MOND effects just start to be important (except for the   peculiar mechanism discussed earlier in the case of a sharp transition\cite{MilSan}). This means that, as a first-order approximation, we can simply consider the density of the ring in General Relativity as an upper limit on the MOND density, and compare it to the Tremaine-Gunn limit. The convergence parameter is $\kappa=0.69$ in the ring \cite{J07}, but the background is estimated \cite{J07} to contribute up to $\kappa=0.65$, which would be the convergence if no ring was present, meaning that the convergence due to the ring itself is $\kappa_r=0.04$. Adopting the effective distance $D_{\rm eff}=D_l D_{ls}/D_s = 0.9$~Gpc (where $D_s$, $D_l$, and $D_{ls}$ are the distance from the observer to the source, from the observer to the lens, and from the lens 
to the source, respectively), we find that the MONDian upper limit of the surface density of the ring is $\Sigma = \kappa_r \times \Sigma_c = 70 {\rm M}_\odot {\rm pc}^{-2}$. Given that the ring is $25''$ wide, i.e. 0.15~Mpc wide for a distance of 1.2~Gpc, it is sensible to consider that its depth along the line-of-sight is of the same order of magnitude leading to $\rho = \Sigma / (0.15$~Mpc)$=4.6 \times 10^{-4} {\rm M}_\odot {\rm pc}^{-3}$, i.e. significantly less (at more than 2$\sigma$) than the Tremaine-Gunn limit. We thus conclude that the ringlike structure in Cl0024+17, if real and not caused by spurious effects in the weak lensing analysis, does not pose a new challenge to MOND in galaxy clusters.

\section{Low temperature X-ray emitting groups}

While we have shown that the widely advertised lensing analysis of the clusters 1E0657-56 and Cl0024+17 do not pose any new challenges to the ``MOND + neutrinos" hypothesis, we show hereafter that low-mass X-ray emitting groups do provide a much more serious problem. Indeed, Eq.~(1) implies that 2~eV neutrinos would stop contributing significantly to the mass density in cooler clusters or groups, since their maximum density is proportional to $T^{3/2}$. The pure ``MOND + neutrinos" hypothesis thus predicts that the MOND mass discrepancy should decrease with decreasing temperature. However, when analyzing the hydrostatic equilibrium of X-ray emitting groups with $0.6 \, {\rm keV} < T < 2 \, {\rm keV}$, in which neutrinos cannot cluster, one finds\cite{AFB} a mass discrepancy that cannot be explained by neutrinos. This of course does not mean that 2~eV neutrinos cannot be present to alleviate the mass discrepancy in rich clusters, but it means that there is {\it more} MOND hidden mass than just neutrinos, especially in cool groups.

\section{Conclusion}

We thus conclude that, while having an amazing predictive power on galactic scales, the simple MOND prescription fails at present in galaxy clusters, where some dark matter is needed.
If this dark matter is assumed to be in the form of 2~eV neutrinos (at the limit of experimental detection), then the bulk of the problem can be solved in rich clusters, including the bullet cluster and the ringlike feature observed in Cl0024+17. However, neutrinos cannot cluster in cool groups with $0.6 \, {\rm keV} < T < 2 \, {\rm keV}$, where a discrepancy is still observed. One solution could then be that dark matter in MOND is in the form of a 4th sterile neutrino with a mass around 6-10~eV. Another possibility is that the new fields that are invoked in relativistic versions of MOND might behave as a dark matter fluid in galaxy clusters\cite{Zhaolambda}.  However these explanations seem to be slightly acts of the last resort, whilst another, more elegant, possibility would be that the MOND cluster dark matter is simply in the 
form of cold gas clouds or MACHO's, since there are enough missing baryons at low redshift to account for all the MOND hidden mass in galaxy groups and clusters (except if more baryons are detected 
in WHIM in between). An interesting possibility is then that this baryonic dark matter is in the form of dense clumps of cold gas of only a Jupiter mass and a temperature of a few Kelvins\cite{PC}, which would behave in a collisionless way.
In any case, one should understand why this MOND dark matter component vanishes for systems with $T<0.6$~keV. As a final remark, it should be highlighted that this additional unseen component in MOND only appears in systems with an abundance of ionised gas and X-ray emission, whatever consequence this might have on the nature of this dark matter.

\end{document}